# Optically Detected Magnetic Resonance Imaging and Sensing Within Functionalized Additively Manufactured Microporous Structures

Brian W. Blankenship[1†], Yoonsoo Rho[2†], Zachary Jones[3,4], Timon Meier[1], Runxuan Li[1], Emanuel Druga[3], Harpreet Singh[5], Xiaoxing Xia[6], Ashok Ajoy[3,7,8*], Costas P. Grigoropoulos[1*]

1. Laser Thermal Laboratory, Department of Mechanical Engineering, University of California, Berkeley, CA 94720, USA
2. Department of Mechanical Engineering, Ulsan National Institute of Science and Technology (UNIST), Ulsan, 44919 Republic of Korea
3. Department of Chemistry, University of California, Berkeley, CA 94720, USA
4. Advanced Biofuels and Bioproducts Process Development Unit (ABPDU), Biological Systems and Engineering Division, Lawrence Berkeley National Laboratory, Berkeley, CA 94720, USA
5. Department of Physics, Guru Nanak Dev University, Amritsar, Punjab 143005, India
6. Materials Engineering Division, Lawrence Livermore National Laboratory, Livermore, CA 94550, USA
7. Chemical Sciences Division, Lawrence Berkeley National Laboratory, Berkeley, CA 94720, USA
8. CIFAR Azrieli Global Scholars Program, 661 University Ave, Toronto, ON M5G 1M1, Canada

† These Authors contributed equally
*Corresponding Author

Email: cgrigoro@berkeley.edu, ashokaj@berkeley.edu

## Abstract

Quantum sensing with nitrogen-vacancy centers in diamond has emerged as a powerful tool for measuring diverse physical parameters, yet the versatility of these measurement approaches is often limited by the achievable layout and dimensionality of bulk-crystal platforms. Here, we demonstrate a versatile approach to creating designer quantum sensors by surface-functionalizing multiphoton lithography microstructures with NV-containing nanodiamonds. We showcase this capability by fabricating a 150 μm x 150 μm x 150 μm triply periodic minimal surface gyroid structure with millions of attached nanodiamonds. We demonstrate a means to volumetrically image these structures using a refractive index matching confocal imaging technique, and extract ODMR spectra from 1.86 μm x 1.86 μm areas of highly concentrated nanodiamonds across a cross section of the gyroid. Furthermore, the high density of sensing elements enables ensemble temperature measurements with sensitivity of 0.548 °K/√Hz at 5 mW excitation power. This approach to creating quantum-enabled microarchitectures opens new possibilities for multimodal sensing in complex three-dimensional environments.

## Introduction

Optically addressable electron spins have emerged as powerful quantum sensors capable of measuring diverse physical parameters, including magnetic[1–3] and electric fields[4], temperature[5], pressure[6], and rotational movement.[7,8] Through optically detected magnetic resonance (ODMR), these miniaturized probes have enabled novel applications across a broad range of contexts, from nanoscale *in-vivo* thermometry of single cells[9,10] to sensing materials' physicochemical properties under extreme temperature and pressure conditions.[6,11] Among various solid-state defects such as color centers in diamond (nitrogen-vacancy (NV$^-$) and silicon-vacancy centers[12]), defects in hexagonal boron nitride (hBN)[13], and molecular systems like pentacene-doped organic crystals[14,15], NV$^-$ centers are ideal for high-precision sensing due to their favorable properties. NV$^-$ centers offer long coherence times at room temperature, demonstrate high sensitivity to external stimuli—particularly magnetic fields—and can be optically initialized and read out at common visible laser wavelengths.[16,17] Moreover, diamond's resistance to chemical degradation and thermal decomposition, along with its biocompatibility, makes NV$^-$ centers suitable in a range of settings, from delicate intercellular environments and harsh chemical environments.

Many NV-based sensing approaches rely on bulk crystalline diamond substrates.[17] Such bulk platforms inherently constrain the sensing volume within a hundred nanometers of the diamond surface, limiting the proximity of sensors to the phenomena of interest. This constraint poses challenges for utilizing ODMR sensing in three-dimensional (3D) environments.

To overcome this, nanodiamonds present an alternative paradigm, allowing for greater flexibility in sensor deployment and enabling closer access to observables of interest.[18,19] The small form factor of nanodiamonds (as low as 10 nm) allows them to be introduced into a variety of environments, including living cells[20,21], fluidic interfaces[22], and complex material interfaces.[23,24] Furthermore, the ability to disperse nanodiamonds within a medium or onto surfaces opens opportunities for integrating NV$^-$ centers into micro-architected materials[25], microfluidic systems[22], and photonic devices[26], enabling 3D sensing across microvolumes or larger scales and expanding the applicability of ODMR-based sensing.

Recent advances have demonstrated the incorporation of nanodiamonds into complex microstructures fabricated via multiphoton lithography (MPL).[25] MPL, a high-resolution 3D printing technique, enables the creation of intricate structures with feature sizes as low as 100 nanometers.[27] By embedding nanodiamonds into a precursor resin for MPL fabrication, it becomes possible to create 3D structures imbued with nanodiamonds containing NV$^-$ centers, combining their sensing capabilities with geometric versatility made available by MPL.[25,26] That said, however, incorporation of nanodiamonds into MPL polymer resin can compromise both feature resolution and quality of MPL, thereby limiting the achievable density of NV$^-$ centers within the microstructures.[25,28] Moreover, optical scattering through complexly structured media makes it challenging to extract and spatially resolve signals from NV$^-$ centers deep within the structures, limiting the effectiveness of ODMR-based sensing.

In this paper, addressing these challenges, we present a new method of functionalizing MPL-fabricated structures with NV$^-$ centers. By immersing MPL-constructed structures in a concentrated nanodiamond suspension, we achieve relatively uniform surface functionalization while preserving the high spatial resolution inherent to MPL fabrication. The efficacy of this method is demonstrated through ODMR measurements on a complex gyroid structure with microporous features, where surface-bound NV-centers enable quantum sensing throughout the intricate porous network. This capability, and the simplicity of its implementation, opens new opportunities for quantum sensing in complex 3D geometries, allowing for the interrogation of physical parameters throughout the volume of micro-architected materials and functional devices.

## Results and Discussion

Complexly structured 3D scaffolding structures are fabricated using MPL onto which nanodiamonds are attached. Specifically, a gyroid triply periodic minimal surface (TPMS) structure is constructed (See **Supplementary Information**). TPMS are mathematically defined high-surface area geometries exhibiting minimal mean curvature while being periodic in three dimensions.[29] Sets of 5 × 5 × 5 lattice structures composed of unit cells, each measuring 30 μm × 30 μm × 30 μm are fabricated using IP-S resin (Nanoscribe) by a customized MPL setup (details provided in the **Materials and Methods** section). The fluorescent properties of IP-S resin enable straightforward visualization of the scaffold matrix during imaging. This resulted in periodic lattices measuring 150 μm ×150 μm × 150 μm which matches the achievable working distance of most commercial high numerical oil immersion objective lenses. The surface area of the gyroid is estimated to be 712500 μm$^2$. To functionalize the surfaces of the porous gyroid structures with nanodiamonds, bare MPL-fabricated scaffolds are immersed in a droplet containing concentrated 100 nm carboxyl-functionalized nanodiamonds suspended in DI water. The droplets are then exposed to diffuse 365 nm ultraviolet (UV) light for approximately 10 minutes. Following UV treatment, the coverslip holding the structures is then briefly heated on a hot plate at 100 °C until the droplet partially evaporates and reforms around the TPP structures. This controlled heating step concentrates the nanodiamond solution which preferentially wets and conforms in and around the gyroid structure's surface. Finally, the wetted structures are rinsed with a solution of 70% isopropyl alcohol, which facilitates the removal of nanodiamonds that are not adhered to the TPP structures. This process likewise leaves the coverslip largely clear of residual nanodiamonds.

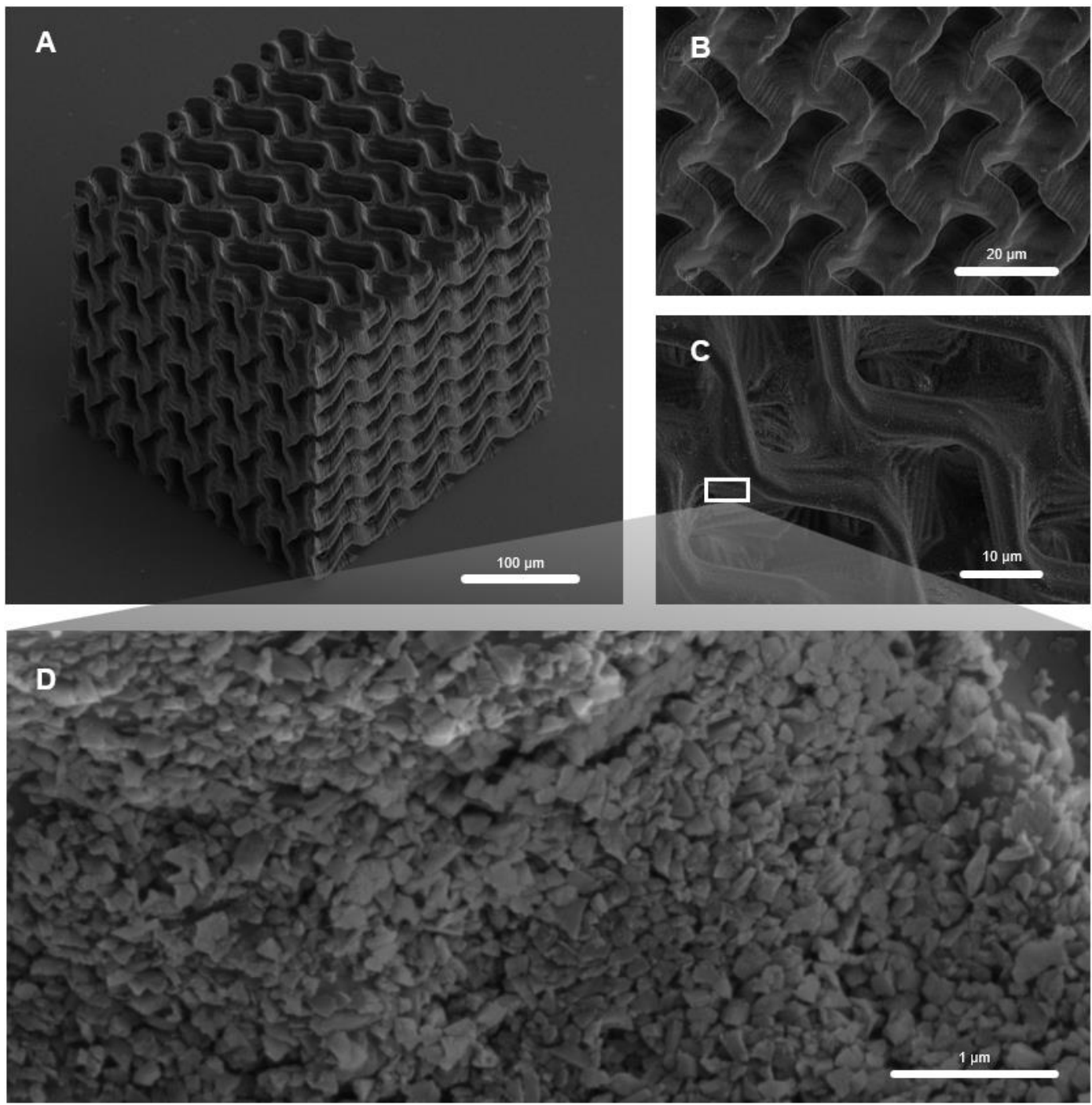

**Figure 1: SEM Images of Nanodiamond-Functionalized Gyroid Structures Fabricated with TPP** (A) Orthogonal view of a 150 μm x 150 μm x 150 μm gyroid structure whose surface is coated with a layer of 100 nm nanodiamonds functionalized with carboxyl groups. (B) Closeup of one of the side faces of the gyroid structure and (C) top surface of the structure where (D) continuous nanodiamond coatings appear to blanket the entire surface of the structure including the surface of interior pores.

Scanning electron microscopy (SEM) images of the nanodiamond-functionalized gyroid structures are presented in **Figures 1A–D**. **Figure 1A** displays an orthogonal view of one of the gyroid scaffolds. The substrate underlying the structure is observed to be relatively clear of nanodiamond buildup. Close-up images of the side and top surfaces are shown in **Figures 1B** and **1C**, respectively, revealing a homogenous coating of nanodiamonds across different facets of the structure. Close inspection of **Figure 1C** reveals that the nanodiamond coating appears to extend into the interior of the pores of the structure. Furthermore, **Figure 1D** provides a highly magnified view of a small section of the top surface, depicting a dense and continuous nanodiamond layer over the complex geometry of the gyroid scaffold. From this image, the density of nanodiamonds is conservatively estimated to be 30 nanodiamonds per $\mu m^2$ assuming a single-layer thick coating and 50% effective surface coverage. This suggest that the structure comprises roughly 2.1 million nanodiamonds- many orders of magnitude more than previously achieved in functionalized MPL structures given that the density is no longer constrained by the manufacturability of the structure.[25,26]

For comparison, a second method consistent with previous MPL methods was employed wherein 100 nm nanodiamonds were directly incorporated into a transparent resin (IP-Visio) prior to dispensing onto the substrate and subsequent laser processing (details provided in the **Supplementary Information**). Although the nanodiamond dispersion appeared relatively uniform before processing, the fabricated structures exhibited signs of clumping of both resin and nanodiamonds, leading to diminished print surface quality. Furthermore, the nanodiamonds were not uniformly distributed across the surface. These results are consistent with previous observations of nanodiamond functionalized MPL structures described in literature.[28]

Laser scanning confocal fluorescence microscopy was used to characterize the distribution of nanodiamonds within the MPL-fabricated gyroid structures. The intricate geometry and inherent scattering properties of these structures posed significant challenges for volumetric imaging in air. To mitigate scattering and enable high-resolution visualization of internal features, the gyroid scaffolds were immersed in a refractive index (RI) matching fluid (RI ≈ 1.518), closely corresponding to that of the polymer material (RI ≈ 1.5). A diagram of this imaging setup is shown in **Figure S4**. Without using an RI matching immersion medium, imaging is restricted to superficial layers due to rapid scattering attenuation.[30,31] The significant refractive index mismatch between the polymer scaffolds and water (RI = 1.33) precludes volumetric imaging in aqueous environments. However, this constraint can be circumvented by employing hydrogel-based scaffolding materials whose refractive indices more closely match that of water.[32] This approach enables ODMR-based sensing in aqueous environments and engineered tissue scaffolds, substantially broadening the potential applications of this technology.

A three-dimensional rendering of the gyroid structure constructed with single-color confocal slices is shown in **Figure 2A**. Representative dual-channel confocal slices at various cross-sectional depths are presented in **Figure 2B-E**. The axial thickness of each confocal slice is expected to be roughly 775 nm. Dual color confocal images are constructed by spectrally separating emissions from the nanodiamond-functionalized structures into two distinct PMT channels based on wavelength. The green channel (520–620 nm) predominantly captures autofluorescence from the IP-S resin, while the red channel (675–775 nm) isolates emissions from negatively charged nitrogen-vacancy ($NV^-$) centers in the nanodiamonds. This spectral filtering effectively suppresses fluorescence from neutral $NV^0$ centers, which have a zero-phonon line at 575 nm, thereby enhancing the specificity for ODMR-active $NV^-$ centers whose zero-phonon line is at 637 nm. Photoluminescence spectra of both the nanodiamonds and the IP-S resin are provided in **Figure S5** for reference. Autofluorescence from the resin decays rapidly as the resin photobleaches, whereas the fluorescence from the nanodiamond regions remains stable with extended exposure to the excitation laser.

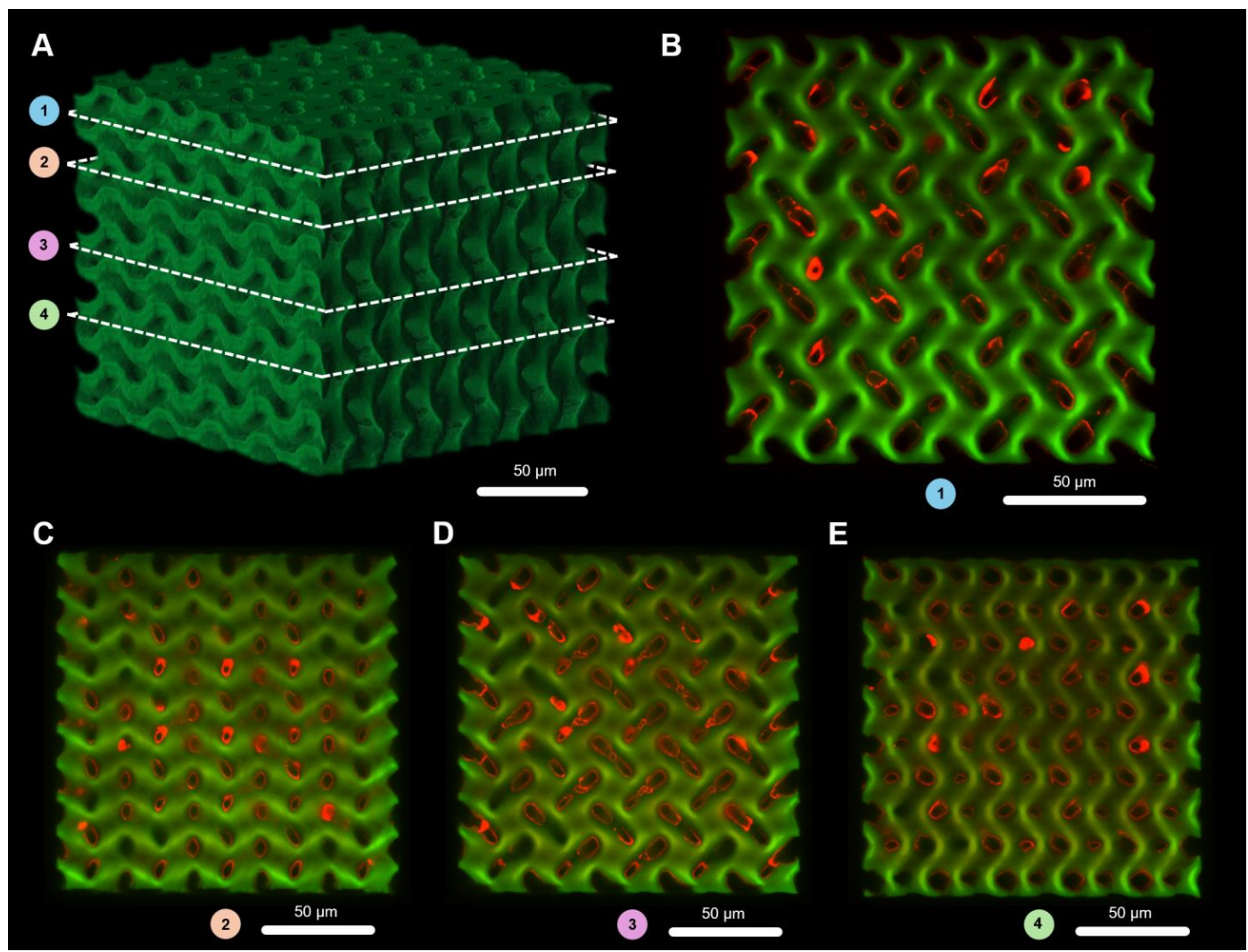

**Figure 2: Optical Reconstruction of Nanodiamond-Coated Structures** (A) Single-color 3D optical reconstruction of the TPP structure with (B-E) four dual-color confocal slices taken at various layers across the structure. Images are generated with 488 nm excitation. Collected fluorescence is spectrally separated into two channels (green 520 - 620 nm) and (red 675 – 775 nm), which makes the red channel more sensitive to NV⁻ fluorescence. Areas in red are representative of locations functionalized with nanodiamonds.

These images resolve the complex architecture of the gyroid with high fidelity and distinctly highlight nanodiamond-functionalized regions. The consistent fluorescence signal along the exterior surfaces confirms a consistent nanodiamond coating throughout the scaffold. Notably, the fluorescence across multiple depths indicate effective attachment not only on the exterior surfaces of the gyroid but also within the internal channels. This enables strategic compartmentalization of sensing regions within the micro-architected channels. In the context of fluidic sensing applications, such three-dimensional functionalization substantially increases the effective interface between target analytes and nanodiamond-modified surfaces, potentially enhancing detection sensitivity.

ODMR offers a non-invasive technique to extract temperature and magnetic field information from NV⁻ centers in diamond.[33] By applying microwave radiation at frequencies resonant with transitions between the spin sub-levels of the NV⁻ center ground state, measurable changes in optical emission intensity are observed. This variation in photoluminescence enables the optical interrogation of energy levels by tuning the microwave frequency. At room temperature, the ground-state spin triplet exhibits a zero-field splitting (ZFS) of approximately $D = 2.87$ GHz.[34]

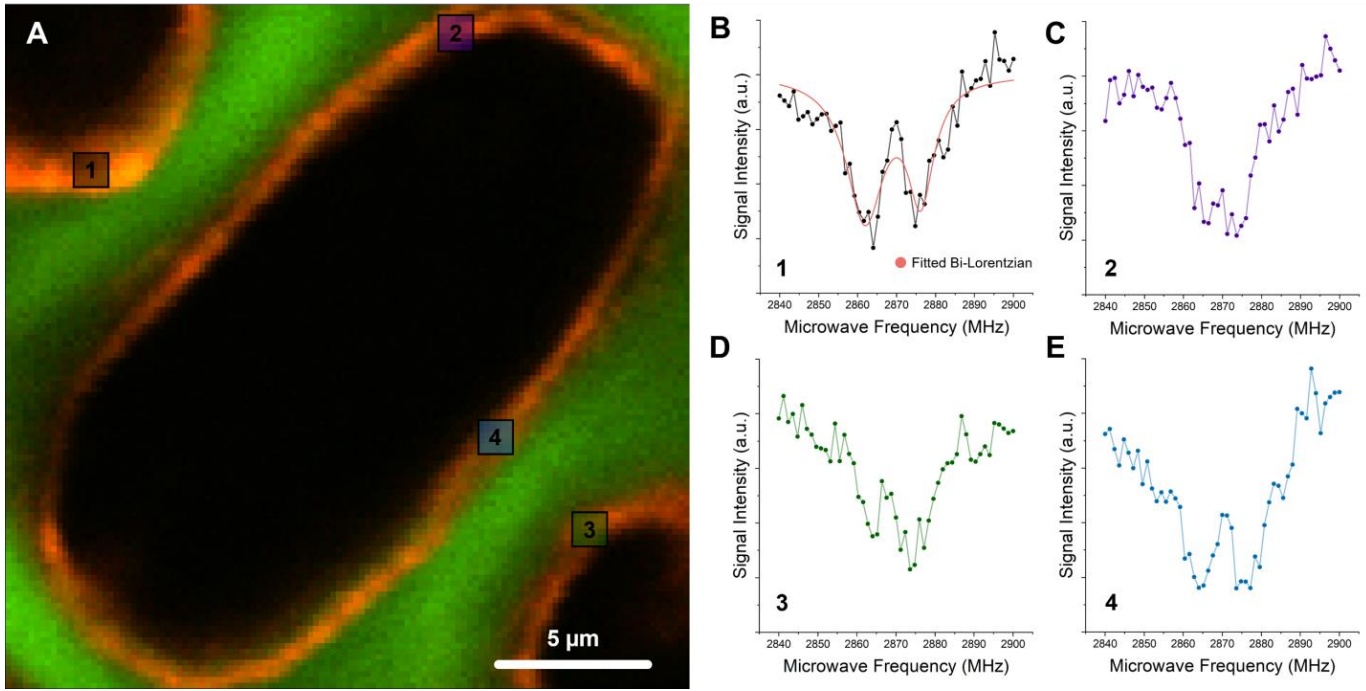

**Figure 3: Spatially Resolved ODMR** (A) Confocal fluorescence image of a printed cross section of the gyroid structure design lined with nanodiamonds. Pixel intensities at selected locations (indicated by markers) were aligned, binned, and plotted as a function of the applied microwave frequency to generate localized ODMR spectra. (B–D) Corresponding ODMR spectra for the selected locations in (A). Bi-Lorentzian are fitted in (B) to outline the characteristic resonance feature of the NV$^-$ center which appears in each curve. The color of each spectrum corresponds to the respective location in (A).

Spatially resolved ODMR measurements are taken at ambient temperature across a printed cross-section of the gyroid structure. Sequential confocal images are acquired by scanning a continuous-wave 488 nm laser beam across the sample. The confocal images were recorded at a resolution of 256 × 256 pixels where each frame synchronized to the application of a monotone microwave frequency delivered via a 4 mm copper coil positioned approximately 2 mm from the sample. The microwave frequency was varied in 1.2 MHz intervals over the range of 2840–2900 MHz. To enhance signal reliability, multiple frames at each frequency were acquired and averaged.

In the selected locations with rich nanodiamond concentrations depicted in **Figure 3A**, pixels were binned in groups of 6 × 6 representing 1.86 µm x 1.86 µm areas to generate distinct ODMR spectra in regions of high nanodiamond concentration, as shown in **Figure 3B–E**. The total acquisition time to capture fluorescence from each 6×6 area at a single microwave frequency was 21.6 ms. ZFS positions of the spectra displayed in **Figures 3B-E** are calculated to be 2870.2, 2869.6, 2867.6, 2870.8 MHz respectively by using a bi-Lorentzian fitting method. It's likely that higher SNR measurements are achievable by enlarging the pinhole aperture to accept a larger percentage of photons, albeit at the cost of increasing the optical sectioning and reducing the spatial resolution.

Extending this methodology to three-dimensional sensing engenders concerns about integration time, as the acquisition duration scales not only as the product of the number of pixels in the *x* and *y* dimensions taken by the confocal microscope, but also the number of *z* slices, as well as the number of microwave frequency steps. To address these challenges and improve sensitivity, four strategies are proposed.

First, utilizing a laser wavelength closer to the resonance frequency of the NV$^-$ centers can improve the fluorescence contrast of NV- centers and improve the SNR of the measurements. Secondly, since the number of NV- centers is expected to scale cubically with the size of the nanodiamonds used, it would be expected that using larger nanodiamonds would likewise improve the SNR. For example, using 150 nm nanodiamonds could increase the signal by approximately 3.375 times due to the higher NV$^-$ count. Third, implementing advanced sampling techniques such as Bayesian sampling can optimize the selection of microwave frequencies, intelligently choosing both the number and specific frequencies applied to maximize sensitivity.[35] Finally, since much of the integration time in image acquisition is spent capturing fluorescence from non-functionalized regions of the microstructure, then by selectively sampling points of interest at the nanodiamond-structure interface, acquisition times can be significantly reduced. Collectively, these implementations promise to improve the necessary integration periods of these measurements by more than an order of magnitude.

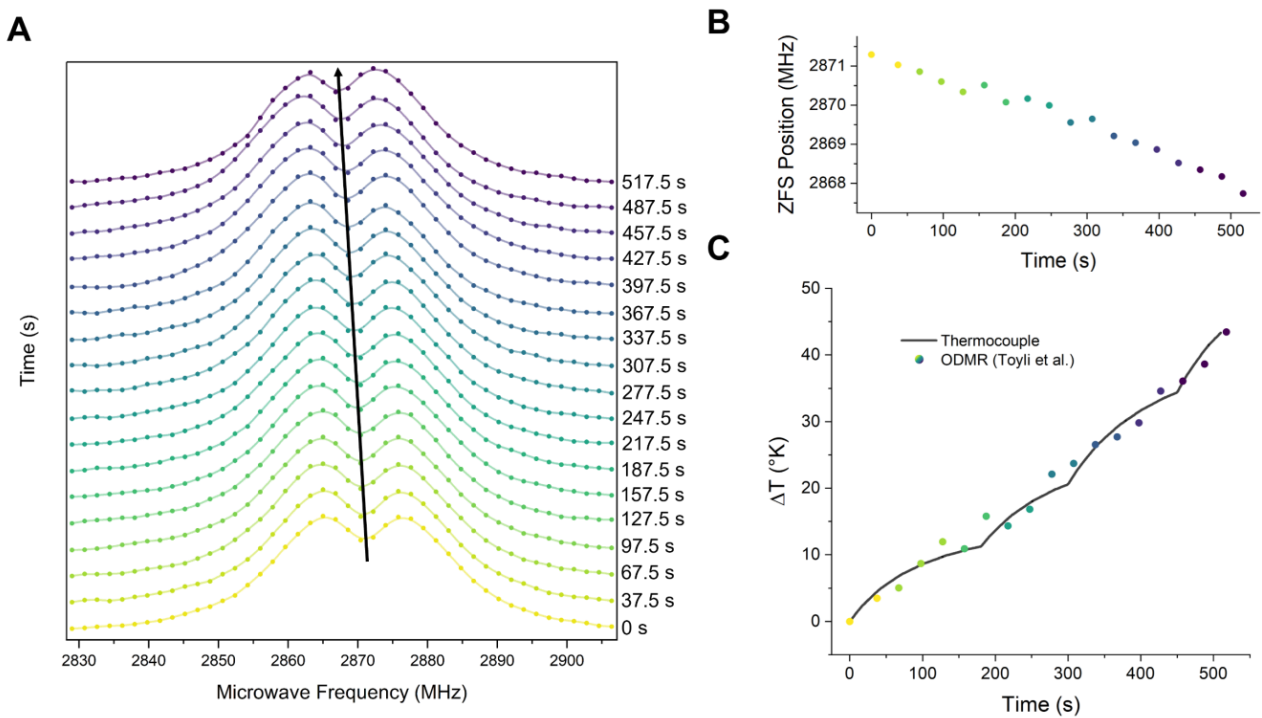

**Figure 4: Temperature Measurements** (A) Waterfall plot of ODMR spectra collected using wide-field microscope. (B) The fitted ZFS position with relation to time which is used to calculate the change in temperature in (C) from the onset of the experiment which is referenced to the change in temperature measured by a nearby thermocouple.

Rather than interrogating discrete points within the gyroid architecture using spatially resolved ODMR, it is also possible to utilize ensemble ODMR to simultaneously probe the conditions across a larger illuminated volume- in this case from millions of nanodiamonds across the gyroid structure. By integrating signals across such a broad region, ensemble ODMR significantly increases signal-to-noise ratios compared to localized confocal imaging, enabling global assessments of conditions using more rapid signal acquisition times. As a temperature sensing demonstration, a coverslip containing a nanodiamond-coated gyroid structure is placed onto a ceramic heating plate and the ODMR spectra are measured at set time intervals. The ceramic plate was heated by applying a variable voltage in discrete steps. To track the relative change in temperature from the onset of this experiment, a type K thermocouple was positioned proximally to the gyroid.

ODMR measurements were taken using a widefield lock-in microscope described previously.[25] The system applies amplitude-modulated microwave pulses synchronized with a lock-in amplifier. This modulation induces a corresponding fluorescence response in the nanodiamonds, oscillating at the same modulation frequency. The lock-in amplifier extracts the fluorescence signal component at this frequency. The structures were excited using a 5 mW, 532 nm continuous-wave Gaussian laser excitation, having a full width at half maximum of approximately 50 μm and focused near the center of a gyroid structure. A 50-point ODMR spectra taken at linearly spaced microwave frequency intervals ranging from 2820 – 2910 MHz at 30 second timesteps using an integration time of 100 ms per point. These spectra are shown in the waterfall plot in **Figure 4A** where the ZFS is observed to shift to lower frequencies in each subsequent spectra over time as the temperature of the structure increases. This shift is illustrated in **Figure 4B.**

The temperature of these ensembles can be determined by analyzing the temperature-dependent ZFS in each spectrum. This temperature dependence originates from thermally induced stress-driven delocalization of the spin-defect wavefunction within the diamond crystal lattice. Following the work of Doherty et al.,[36] the splitting parameter $D$ can be expressed as:

$$D \propto C\eta^2 \left\langle \frac{1}{r^3} - \frac{3z^2}{r^5} \right\rangle \qquad (1)$$

where $C$ represents the spin-spin interaction constant, $\eta$ denotes the electron density, and $\left\langle \frac{1}{r^3} - \frac{3z^2}{r^5} \right\rangle$ characterizes the interaction of sp$^3$ electron densities in carbon atoms. The splitting parameter exhibits a non-linear variation with temperature, arising from thermal expansion of the diamond lattice that increases lattice spacing $r$. This phenomenon manifests as a frequency shift towards lower values at an approximately linear rate of -74 kHz/°K near room temperature.[37,38]

Zero-field splitting for each spectral curve is determined through bi-Lorentzian curve fitting of ODMR spectra, with relative temperature changes calculated using the relations established by Toyli et al.[37] These measurements are compared against corresponding thermocouple recordings, with the comparative temperature variations illustrated in **Figure 4C**. The Biot number of the fabricated gyroid structure is estimated to be << .1, which suggests that it should maintain a fairly uniform temperature distribution across its volume. An estimation of the temperature sensitivity is made using the equation:

$$\text{Sensitivity} = \frac{\sigma_P \sqrt{\delta_{int}}}{dS/dT} \qquad (2)$$

Where, $\delta_{int}$ is the point integration time, $\frac{dS}{dT}$ is the maximum ODMR signal slope with respect to temperature and $\sigma_P$ is the signal noise calculated as the standard deviation of the signal local to the point of maximum slope. The average temperature sensitivity of these measurements is calculated to be 0.548 °K/√Hz. These measurements are found to be more sensitive than previous measurements taken on MPL structures fabricated with nanodiamond-containing resin despite using 20 times lower optical excitation power and comparable microwave powers. This can be attributed to the increased number of nanodiamonds within the sensing region achieved through surface functionalization, compared to bulk incorporation methods previously reported.[25] However, the sensitivity of these measurements may still be limited by thermal drift during finite acquisition times.

MPL affords an exceptional level of design flexibility in fabricating microscale structures precisely tailored to experimental or application-specific requirements. When functionalized with nanodiamonds, these can enable non-invasive, spatially and temporally resolved measurements of parameters such as temperature gradients, magnetic fields, and paramagnetic ion concentrations. In the context of thermometry, such NV-functionalized devices can offer real-time insights into reaction and diffusion processes in porous microenvironments by capturing three-dimensional temperature gradients throughout a microvolume.

Furthermore, there are enticing opportunities for applications that enable local and time-resolved in-situ detection of chemical constituents and reactions. Applications that require on local enhancement predominantly rely on surface-enhanced infrared (SEIRAS) or Raman spectroscopy (SERS), which use specific metallic catalysts (Au, Ag, Pd), thereby imposing significant analytical constraints.[39] Conversely, many bulk characterization tools lack the spatial resolution often required or *ex-situ* techniques like mass spectrometry present challenges for time-resolved measurements. In these contexts, functionalized MPL structures offer a promising alternative for non-invasive, real-time detection, building upon previous spin-based measurements of paramagnetic ion interactions at lower NV$^-$ center densities.[40] Further functionalized MPL structures may be additionally equipped with surface-anchored reactive centers such as enzymes or catalysts allowing direct spectroscopic insights into these entities. Such capabilities can potentially expand the diagnostic repertoire available in microscale fluidic environments and offer a means to explore new insights into environments that were once thought too challenging to probe.

## Conclusion

This study introduces a new approach to creating designer quantum sensors by surface functionalizing MPL microstructures with nanodiamonds. Such method can create millimeter-sized 3D architectures with feature sizes below 1 μm with dense coatings nanodiamonds without compromising structural integrity or print quality. These functionalized micro-architected scaffolds offer a framework for multimodal, multidimensional sensing using optically detected magnetic resonance from $NV^-$ centers. As a proof of concept, millions of nanodiamonds are attached to the complex surface of a 150 μm × 150 μm × 150 μm TPMS gyroid lattice structure with porous interior. Spatially resolved quantum sensing using ODMR is demonstrated at lateral resolutions of 1.86 μm at select locations of high nanodiamond concentration within the interior of the gyroid structure. This is achieved by using confocal fluorescence microscopy technique that employs refractive index matching fluids and is synchronized with microwave frequency application. Moreover, time-resolved ensemble temperature sensing is demonstrated with sensitivities as low as 0.548 °K/$\sqrt{Hz}$ using 5 mW 532 nm CW excitation. This methodology promises to enable new opportunities for application-specific multimodal sensing in complex fluidic environments.

## Materials and Methods

Multi-Photon Lithography: A custom-built multi-photon lithography system is used to fabricate three-dimensional gyroid structures as scaffolding for nanodiamonds. The system tightly focuses a near-infrared femtosecond laser beam (InSight®X3+, Spectra-Physics with $\lambda_{peak}$ ~800 nm, pulse width 100 fs, and repetition rate 80 MHz) inside of the resin by employing an oil immersion 25X/NA0.8 objective lens (Carl Zeiss). The system can form a voxel size with lateral 500 nm and axial 3.46 μm resolutions in the Abbe diffraction limit condition inside the printing resin with a refractive index 1.5 (IP-s, Nanoscribe). The high photon flux (~1 TW/cm$^2$) produced by the tightly focused femtosecond pulse induces spatially confined cross-linking of resin through two-photon polymerization process. A two-axis galvanometer mirror scanner steers the femtosecond laser beam, and coupled with 4f system, it translates the voxel in a lateral direction at 100 mm/s scan speed with 0.25 μm hatching. A linear z-stage changes the distance between the sample and the objective lens in a vertical direction by 1 μm step after finishing each layer. After finishing printing, the printed structure is immersed in a developing bath of propylene glycol monomethyl ether acetate (PGMEA) for 1 hour, followed by dipping in isopropanol for another 1 hour to remove any remaining uncross-linked resin and solvent.

Nanodiamonds were concentrated by centrifuging a solution of 0.1% (w/v) 100nm nanodiamond (Adamas Nanotechnologies) on a benchtop centrifuge at 2500g relative centrifugal force for 30 minutes. The concentrated solution of nanodiamonds was extracted by inserting a micropipette into the sediment. This concentrated solution was then used for coating the MPL structures.

Confocal Imaging: An Olympus Fluoview 3000 laser scanning microscope was used for single-photon imaging using an excitation wavelength of 488 nm and using a 488 nm long pass filter. Collected fluorescence was split into two channels via a spectral detection unit in line with two detectors. Images were generated of the structure at a variable pixel resolution and 12-bit intensity resolution. A 60X, 1.35 NA oil immersion objective (Olympus) was used, and the pinhole aperture was set to 1 Airy unit.

Photoluminescence Spectra: Photoluminescence measurements were taken with a Renishaw InVia spectrometer with 1800 lines/mm grating.

Point ODMR measurements were taken using a custom built wide-field lock-in fluorescence microscope with 20X objective discussed in detail in previous works.[25]

Image Processing: The gyroid structures exhibited translational shifts within the microscope's field of view during continuous imaging, primarily due to thermally induced movements arising from non-uniform heat flux associated with microwave frequency changes. To mitigate this lateral drift, we implemented a standard scale-invariant feature transform (SIFT) algorithm during post-processing, allowing for computational realignment of the structures relative to an initial reference frame.[41]

Temperature control: MPL structures were placed on a ceramic heating element. A variable voltage power supply was hooked up to the heating element and the voltage was raised in discrete 1.5 V increments at t = 180s, t = 300s, and t = 450 s. A Keithley DAQ6510 with type K thermocouples was used to track the temperture across the duration of the experiment.

## Supplementary Information

Geometry of gyroid structure, images of nanodiamond-containing resin, SEM images of alternative MPL fabrication technique using nanodiamond-containing resin, illustration of the imaging setup, photoluminescence measurements of resin and nanodiamonds, bi-Lorentzian fitting parameters, temperature relations, fitted ODMR curves.

## Acknowledgements

Brian Blankenship acknowledges support from the NSF Graduate Research Fellowship (DGE 2146752). We acknowledge support from DOE BER (DESC0023065), William M. Keck Foundation (8959), DOE SBIR (DE-SC0022441), AFOSR YIP (FA9550-23-1-0106), and the CIFAR Azrieli Foundation (GS23-013). This work was partially funded by the Laboratory Directed Research & Development program at the E. O. Lawrence Berkeley National Laboratory. Xiaoxing Xia and Yoonsoo Rho acknowledge support by Lawrence Livermore National Laboratory's Lab Directed Research and Development Program (23-ERD-027). Work at LLNL was performed under the auspices of the U.S. Department of Energy by Lawrence Livermore National Laboratory under Contract DE-AC52-07NA27344. Brian Blankenship acknowledges Dr. Arik Beck and Dr. Andrea Blankenship for fruitful discussions on potential applications. SEM images were taken with the Scios 2 DualBeam available at the Biomolecular Nanotechnology Center of the California Institute for Quantitative Biosciences (QB3), UC Berkeley.

**Gyroid Geometry**

The gyroid structure presented in Figure 1 was created using TPMS Designer software.[1] Its surface is described by a triply periodic minimal surface (TPMS) function:

$$Gyroid, U = \cos 2\pi x \sin 2\pi y + \cos 2\pi y \sin 2\pi z + \cos 2\pi z \sin 2\pi x$$

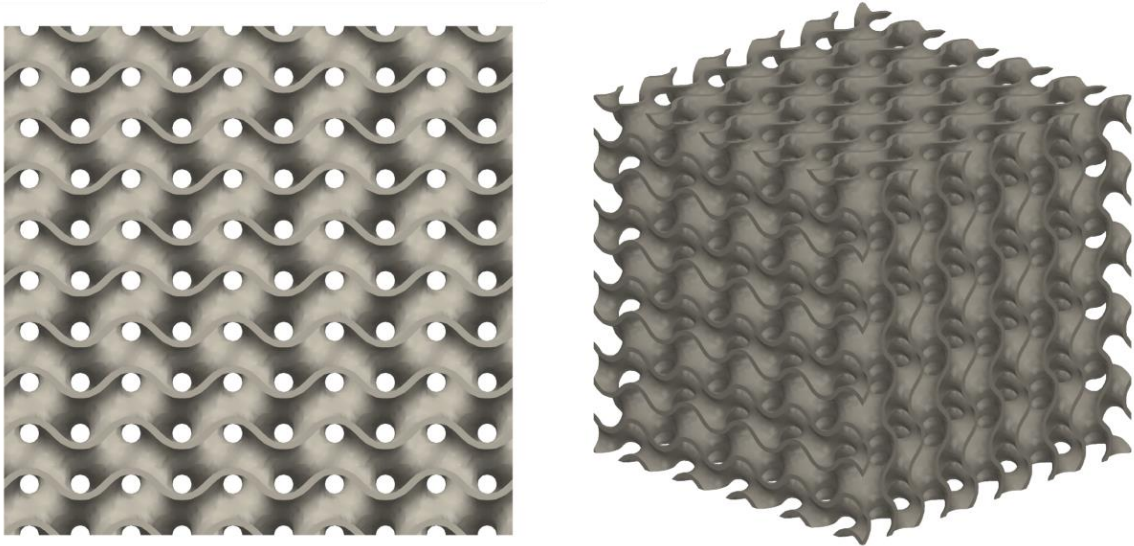

**Figure S1:** (Left) Front and (Right) Orthogonal view of the target gyroid geometry as generated in TPMS Design software. The resulting triply periodic minimal surface forms a continuous, porous network that allows fluid to flow freely through its interior.

The calculated volume fraction was .2486. The fabricated gyroid structure is a 5 x 5 x 5 lattice composed of 30 μm × 30 μm × 30 μm unit cells. A .STL file of this geometry was generated and used as a target geometry for MPL fabrication.

**Nanodiamond-Containing Resin**

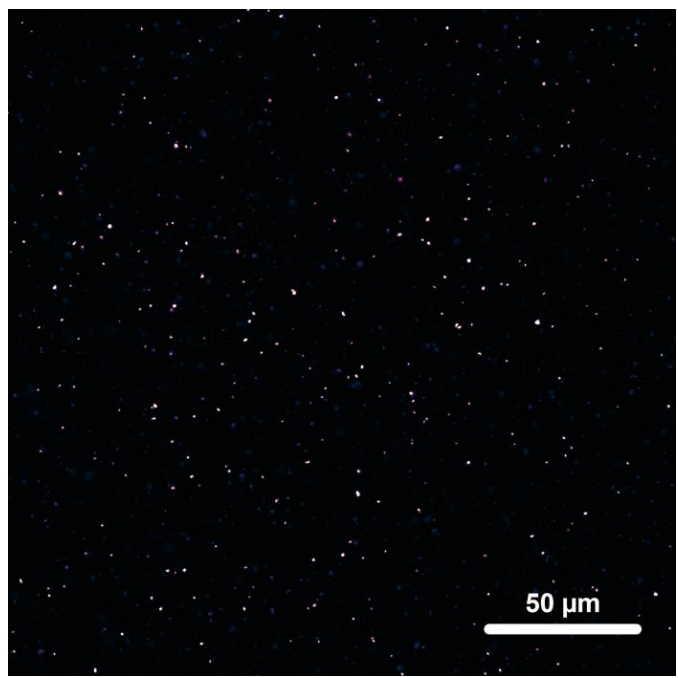

**Figure S2:** 1024 × 1024-pixel confocal fluorescence image of a layer of nanodiamond-containing IP-Visio resin before fabrication, acquired using a 60×, 1.35 NA objective. The field of view is 212 × 212 µm. Unlike IP-S resin, IP-Visio is largely non-fluorescent, allowing the nanodiamonds to be clearly observed within the resin. The pink colorations correspond to fluorescence intensity in the 600-700 nm spectral range, originating from dispersed nanodiamonds.

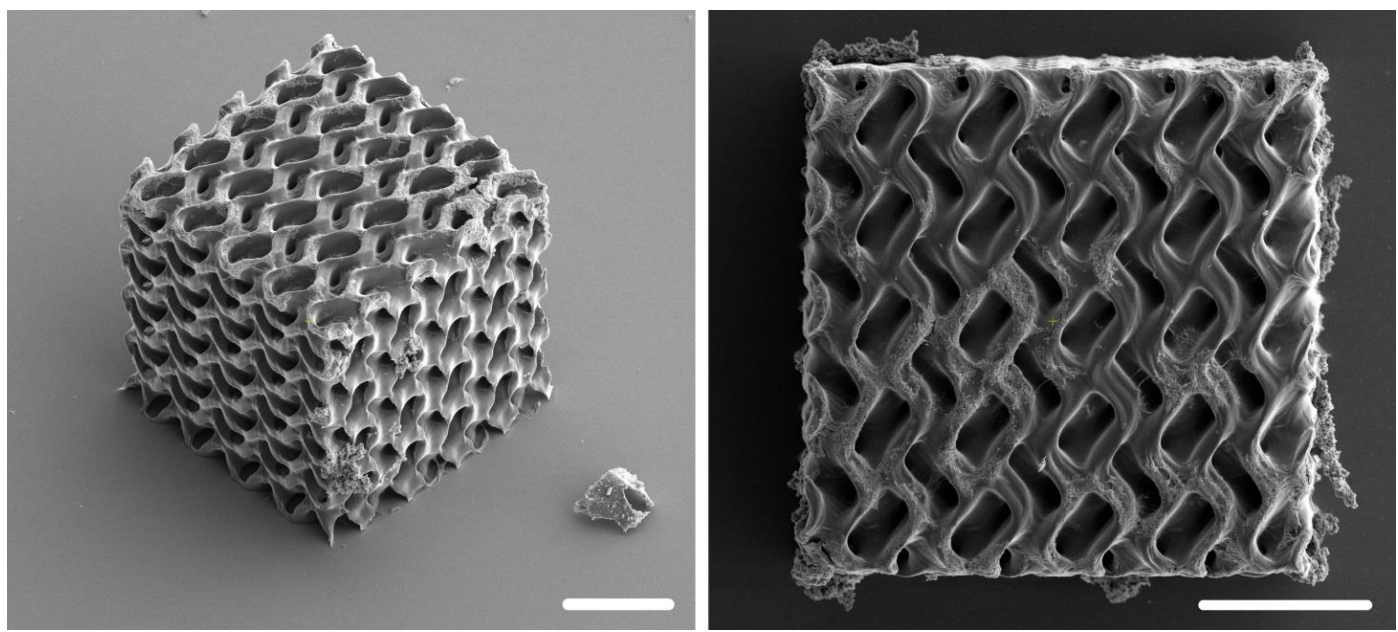

**Figure S3:** SEM images gyroid structure fabricated using resin containing nanodiamonds (Left) Orthogonal view of structures fabricated with the nanodiamond infused resin alongside (Right) closeups of the and top surface. The crust-like coating on the structures appear to contain nanodiamonds. It is likely that particle agglomeration during the dip-in fabrication process contribute to the poor print quality. Scale bars 50 µm.

# Imaging Setup Using Refractive Index Matching Fluid

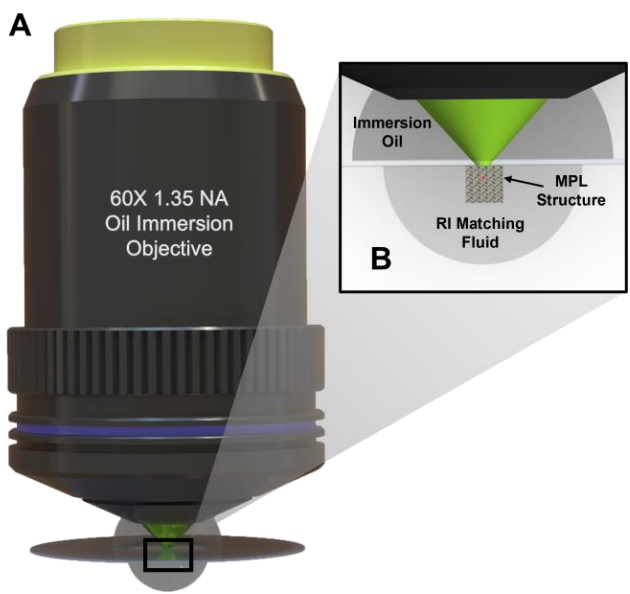

**Figure S4:** (A) Confocal images of the gyroid structure displayed in **Figure 2** are imaged with an Olympus 60X 1.35 NA oil immersion objective on a FluoView 3000 confocal microscope where (B) the nanodiamond-coated MPL structures are enveloped in a droplet of RI-matching oil, which prevents optical scattering an enables imaging that can resolve the interior of the porous structures.

# Photoluminescence Measurements

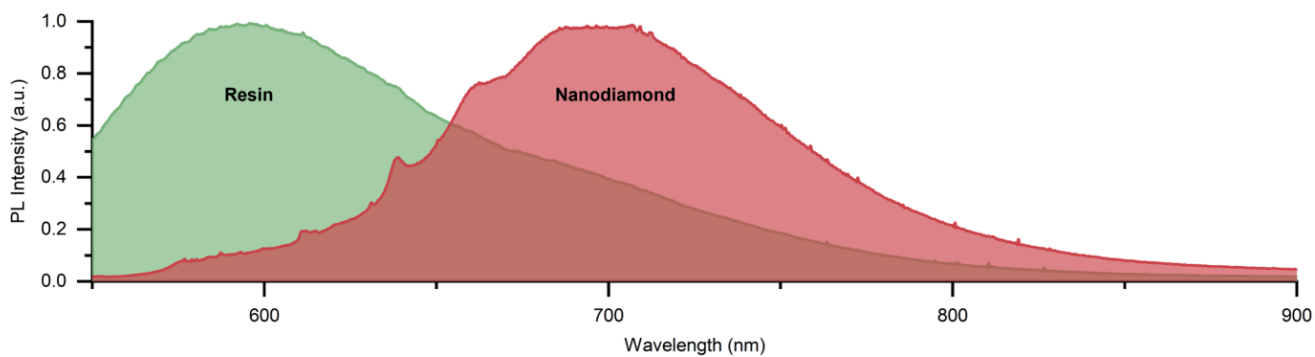

**Figure S5:** PL spectra taken at 532nm excitation of IP-S resin (green) and 100 nm nanodiamonds (red).

**Bi-Lorentzian Fitting**

ODMR curves were fitted to a bi-Lorentzian curves of form:

$$y(x) = \frac{A_1}{[1 + ((x - x_1)/\Gamma_1)^2]} + \frac{A_2}{[1 + ((x - x_2)/\Gamma_2)^2]}$$

Where the fitting parameters are:
$A_1$ and $A_2$ are the amplitudes of the two peaks
$X_1$ and $x_2$ are the center positions of the two peaks
$\Gamma_1$ and $\Gamma_2$ are the half-width and half maximum of the two peaks

**Temperature Relations**

Temperature measurements in Figure 4 were calculated by relating the fitted ZFS to the following relation from Toyli et. al:[2]

$$D(T) = a_0 + a_1 T + a_2 T^2 + a_3 T^3$$

Where:

$a_0 = 2869.7$ MHz
$a_1 = 9.7 * 10^{-2}$ MHz/°K
$a_2 = -3.7 * 10^{-4}$ MHz/°K$^2$
$a_3 = 1.7 * 10^{-7}$ MHz/°K$^3$

## Normalized ODMR Curves

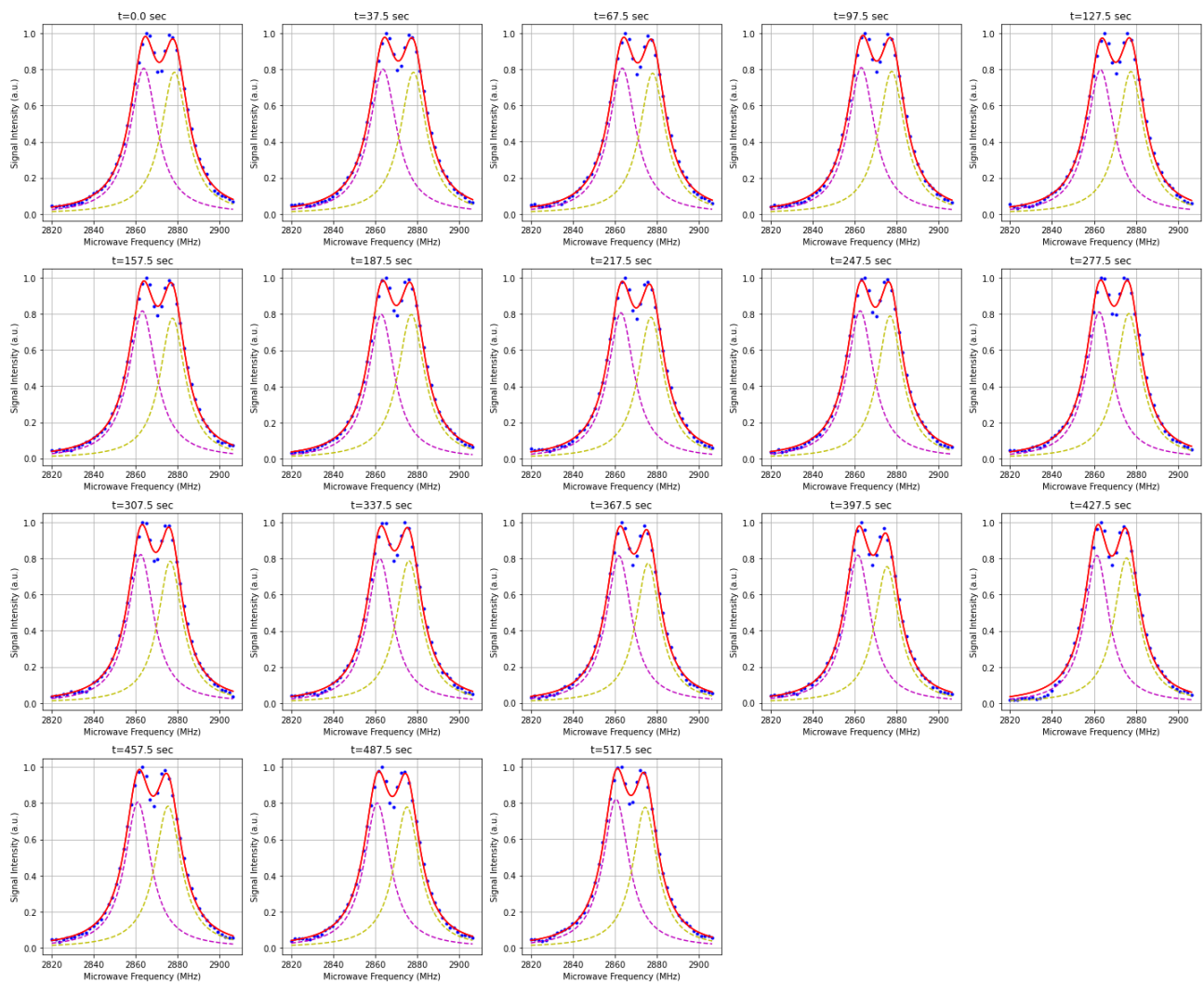

**Figure S6:** Fitted bi-Lorentzian curves to normalized transient ODMR data.